\documentclass[12pt,onecolumn,peerreview,letterpaper]{IEEEtran}
\usepackage{amsmath,amssymb,amsfonts,enumerate,amsthm,graphicx}
\usepackage{dsfont}
\usepackage{array}
\usepackage{cite}
\usepackage{float}
\usepackage[T1]{fontenc}
\usepackage{array}
\usepackage[pdftex,bookmarks]{hyperref}
\usepackage{subfig}

\newtheorem{definition}{Definition}

\newtheorem{example}{Example}
\begin{document}
\title{LLR Approximation for Wireless Channels Based on Taylor Series and Its Application
to BICM with LDPC Codes}

\author{Reza ~Asvadi\authorrefmark{1},
        Amir ~H. Banihashemi\authorrefmark{2},~\IEEEmembership{Senior~Member,~IEEE,}
        \\ Mahmoud ~Ahmadian-Attari\authorrefmark{1}, and Hamid
        ~Saeedi\authorrefmark{3},~\IEEEmembership{~Member,~IEEE}

\authorblockA{\authorrefmark{1}Department of Electrical and Computer Engineering\\
K. N. ~Toosi University of Technology, Tehran, Iran\\
 Emails: asvadi@ee.kntu.ac.ir, mahmoud@eetd.kntu.ac.ir}\\
\authorblockN{\authorrefmark{2}Department of Systems and Computer Engineering, Carleton
University,\\ Ottawa, ON K1S 5B6, Canada\\
 Email: ahashemi@sce.carleton.ca}\\
 \authorblockN{\authorrefmark{3}Department of Electrical and Computer
 Engineering,\\ Tarbiat Modares University, Tehran, Iran\\
 Email: hsaeedi@ieee.org}}
\thanks{A preliminary version of part of this paper was presented at {\em IEEE Globecom 2010},
Miami, Florida, Dec. 6 - 10, 2010.}

\maketitle

\begin{abstract}
A new approach for the approximation of the channel log-likelihood ratio (LLR)
for wireless channels based on Taylor series is proposed.
The approximation is applied to the uncorrelated flat Rayleigh fading
channel with unknown channel state information at the receiver.
It is shown that the proposed approximation greatly simplifies the
calculation of channel LLRs, and yet provides results almost identical to
those based on the exact calculation of channel LLRs.
The results are obtained in the context of bit-interleaved coded modulation
(BICM) schemes with low-density parity-check (LDPC) codes, and
include threshold calculations and error rate performance of finite-length codes.
Compared to the existing approximations, the proposed method is either significantly
less complex, or considerably more accurate.
\end{abstract}

\newpage
\begin{keywords}
Log-likelihood ratio (LLR), LLR approximation, wireless channels, flat fading, Rayleigh fading, uncorrelated Rayliegh fading,
bit-interleaved coded modulation (BICM), low-density parity-check (LDPC) codes, iterative decoding,
decoding threshold, SNR threshold.
\end{keywords}

%\IEEEpeerreviewmaketitle

\section{INTRODUCTION}
In binary transmission over a wireless channel, the derivation of
channel log-likelihood ratios (LLR) is often needed at the receiver
for the detection and/or decoding of information. The channel
LLR is sometimes referred to as {\em soft information}, and its availability
can improve the performance of the detection/decoding schemes significantly.
The channel LLR values depend on the channel output, the noise power and the fading
characteristics as well as the amount of channel state information
(CSI) available at the receiver. In practical systems, however, acquiring the
CSI would require extra bandwidth for the transmission of pilot symbols
and extra complexity at the receiver for the channel estimation. In certain scenarios, this
may not be desirable. It is thus important to derive the LLR
at the absence of CSI. In particular, low-complexity approximations
of LLR are of great practical importance. One should note that
even if efforts are made to estimate the channel and
to attain the CSI, there always exist errors in the estimation process, which
in turn results in imperfect CSI at the receiver. There has been
thus literature on the study of the effect of imperfect CSI on the performance
of transmission schemes (see, e.g., \cite{SB-07}), and on the design
of schemes which are robust to such imperfections (see, e.g., \cite{SB-09}).

At the absence of CSI at the receiver, the
relationship between the channel LLR, $L$, on one hand, and the
channel output $Y$, and the noise power $\sigma_n^2$, on the other
hand, is complex. This can significantly increase the complexity of
a detection/decoding process which relies on the calculation of
channel LLR values. In addition, the complex relationship
$L(Y,\sigma_n^2)$, can impede the analysis and the design of
transmission schemes over wireless channels which depend on the
calculation of the probability density function (pdf) of $L$ as the
starting point. One example is the application of
{\em density evolution}~\cite{rich01B} to the analysis
and the design of coded schemes. Motivated by these, much research has been devoted
to the approximations of channel LLR values for wireless channels;
see, e.g.,~\cite{hou},~\cite{TB-02},~\cite{YA-09},~\cite{YA-10} and the references therein. In
particular, linear or piece-wise linear approximations of $L$ as a function of $Y$ have
received special attention due to their utmost simplicity.

To analyze and design binary low-density parity-check (LDPC) codes
for binary phase shift keying (BPSK) transmission over flat Rayleigh
fading channels in the absence of CSI,
Hou {\em et al.}~\cite{hou} proposed the following linear
approximation of the channel LLR:
\begin{equation}
\hat{L}_A = \frac{2}{\sigma_n^2} E(A) Y \stackrel{\Delta}{=} \alpha_A Y\:,
\label{eqsa}
\end{equation}
where $E(A)$ is the expected value of the channel gain. Though very
simple, this approximation is not very accurate and performs rather
poorly compared to true LLR values~\cite{YA-09}.
%Moreover, for non-binary modulations this method is not simple
%anymore and can cause few tenth of dB performance degradation.
A more accurate linear approximation of LLR was recently proposed in~\cite{YA-09} as
\begin{equation}
\hat{L}_{\hat{C}} = \alpha_{\hat{C}} Y\:,
\label{eqtl}
\end{equation}
where
\begin{equation}
\alpha_{\hat{C}} = \text{arg}\max_{\alpha}\{1-\int_{-\infty}^{\infty}\;\log_2(1+e^{-\hat{l}})f_{\hat{L}}(\hat{l})\;d\hat{l}\}\:.
\label{eqoj}
\end{equation}
In (\ref{eqoj}), $f_{\hat{L}}(\hat{l})$ is the pdf of the
approximate linear LLR parameterized by $\alpha$ according to
$\hat{L} = \alpha Y$. It is shown in~\cite{YA-09} that the
approximation in (\ref{eqtl}) provides considerable improvement
compared to (\ref{eqsa}) for the BPSK modulation and results in
performances very close to those of the true LLR calculation.
The calculation of
$\alpha_{\hat{C}}$ is however, much more complex than that of
$\alpha_A$ and requires solving the convex optimization problem of
(\ref{eqoj}) using numerical techniques.

The general approach of~\cite{YA-09} based on the formulation of (\ref{eqtl}) and
(\ref{eqoj}) was then generalized to non-binary modulation schemes in~\cite{YA-10},
where piece-wise linear approximations of LLR for $8$-PAM and $16$-QAM constellations
were derived. These results were then used to evaluate the performance
of LDPC-coded BICM schemes. It was demonstrated~\cite{YA-10} that for the tested LDPC codes,
the gap in the performance of $8$-PAM for true LLR calculations and the
approximations is rather small (a few hundredths of a dB). This gap however,
is larger for $16$-QAM (about $0.2$ dB).

In this paper, we propose to use the Taylor series of the channel
LLR as the method of approximation. Although many of our results
are in principle applicable to a variety of fading channel models,
in this work, we only consider the uncorrelated flat Rayleigh
fading channel. For the BPSK transmission over this channel with no CSI,
we derive the Taylor series analytically and demonstrate that by using only the first term of
the series, one can obtain an analytical linear approximation which
is almost as simple as (\ref{eqsa}) and yet is practically as
accurate as (\ref{eqtl}). By using the first two terms of the series,
we derive a more accurate analytical non-linear approximation of the channel LLR and
obtain performance improvements compared to the approximation of~(\ref{eqtl}).

For non-binary modulations, we derive piece-wise linear and non-linear approximations
of channel LLRs based on the Taylor series. Compared to the piece-wise linear approximation
of~\cite{YA-10}, our approach is simpler, both conceptually and complexity-wise.
Moreover, our approach can be easily extended from piece-wise linear to non-linear approximations.
This however, may not be simply achievable for the approach of~\cite{YA-10},
where the linearity has an important consequence of making the optimization
problem convex and thus tractable. Performance-wise, we demonstrate that
for the $8$-PAM constellation in an LDPC-coded BICM scheme, our piece-wise
linear approximation performs as good as the approximation of~\cite{YA-10} and very close to
true LLR calculations. We expect this to be the case also for other one-dimensional
constellations. For two-dimensional constellations such as $16$-QAM however,
our second order Taylor series approximation of LLR outperforms the approximation of~\cite{YA-10}
handily, and still performs very close to true LLR calculations.

It is important to note that the complexity of calculating the approximate LLR
is particularly important for two-dimensional constellations. While one might argue that
such computations can be performed off-line and the results can be stored at
look-up tables for different values of noise power and received signal values,
such tables will have to be three dimensional for two-dimensional constellations,
thus requiring much larger storage. If storage is constrained, one may have to
perform the calculations on-line. Another point worth emphasizing is the
importance of the computational complexity of finding the pdf of the LLR (or its approximations) in
the process of analyzing or designing a transmission
scheme using techniques such as density evolution~\cite{rich01B}. These techniques,
which are used for iterative coding/processing schemes (also known as {\em message-passing}
schemes), are based on tracking the pdf of messages throughout iterations, starting from the
pdf of the channel LLR. This is usually performed multiple times for different
values of the channel signal-to-noise ratio (SNR) to find the infimum value of SNR for which
the algorithm succeeds (probability of error tends to zero as the number of iterations
tends to infinity).
This infimum value of the channel SNR is called the {\em threshold}.
One example of such analysis, is to find the threshold of an ensemble of LDPC codes~\cite{rich01B}.
One thus needs to find the pdf of the channel LLR
many times for different SNR values in this process. The number of times such
computations have to be repeated increases even further (by a significant
margin) in a design process. Such a process is usually
based on iteratively optimizing different variables to achieve the best performance.
For example, in the design of irregular LDPC code ensembles, degree distributions of variable
nodes and check nodes are optimized to achieve the best threshold~\cite{RSU-01}.
This design process usually includes an analysis loop which is repeated
numerous times as the design variables are modified to converge to a local
optimum.

The remainder of the paper is organized as follows: Section
\ref{sec::model} is devoted to the fading channel model, LDPC codes,
BICM scheme and the derivation of channel LLRs. In Section \ref{sec::llr}, we
present the Taylor series approximations of the channel LLR for
uncorrelated flat Rayleigh fading channels.
Simulation results are presented in Section \ref{sec::simul},
and finally Section \ref{sec::conclusion} concludes the paper.

\section{CHANNEL MODEL, BICM SCHEME, AND LDPC CODES}
\label{sec::model}
\subsection{Channel Model, BICM and LLR Approximations}
\label{subsec2.1}
Consider the following model of a flat fading channel:
\begin{equation}
\label{eq1}
Y_t = A_t\:X_t+Z_t,
\end{equation}
where $X_t$ and $Y_t$ represent the channel input and output
at time $t$, respectively; $Z_t$ is the zero-mean (possibly complex) Gaussian noise with variance $\sigma^2$ ($2\sigma^2$ for two-dimensional constellations), and $A_t \geq 0$ is the channel gain,
both at time $t$. In this work, we assume that
$A_t$ has a normalized Rayleigh distribution, i.e., $p_{A_t}(a)= 2a\: e^{-a^2}$,
where $p_{A_t}(a)$ is the pdf of $A_t$. We further
assume that sequences $\{X_t\}$, $\{A_t\}$ and $\{Z_t\}$ consist of independent and identically distributed (i.i.d)
random variables. Moreover the three sequences are assumed to be independent of each other.
This model is referred to as the \emph{uncorrelated flat Rayleigh fading channel}.

At the transmitter, the information bit sequence is first mapped to the coded bit sequence
by being passed through the LDPC encoder. The coded bit sequence is then partitioned into blocks
of length $m$. Each block $\mathbf{b}_k=\{b_k^1,b_k^2,\ldots,b_k^m\}$ is then mapped
to a signal $X_k, \:k=1,\ldots,M$, from a Gray-labeled $M$-ary signal constellation
$\chi$ with $M=2^m$ signals.
The block $\mathbf{b}_k$ is referred to as the $k$th {\em symbol} corresponding to the
$k$th signal, and $b_k^i$ is the $i$th bit of the $k$th symbol (signal).
Between the encoder and the modulator, an interleaver conventionally exists.
This however may not be required for the LDPC codes, as the interleaver is
inherent in the structure of these codes when the parity-check matrix is constructed
randomly~\cite{MB-06}.
For $M>2$, this setting is also known as bit-interleaved coded modulation (BICM).
In this work, we assume {\em ideal} interleaving of the bits, which implies that
the transmission of each symbol over the channel is equivalent to the
transmission of its constituent bits over $m$ parallel and
independent memoryless binary-input channels.
These channels are referred to as {\em bit-channels}. At the receiver,
the LLR for each bit-channel is independently calculated.

In the channel model described above, assuming that the noise variance is known at the receiver,
we will have two scenarios depending on the availability of $A_t$ at the receiver:\footnote{In
the rest of the paper, the time index $t$ may be dropped since the distribution of
random variables does not depend on $t$. Also, upper case and lower case variables are used to denote
random variables and their values, respectively, e.g., random variable $Y$ can take
the value $y$.}

\begin{enumerate}
\item \emph{Known CSI}: In this case, the
channel gain $A_t$ is known at the receiver for every $t$. Thus the
channel LLR of the $i$th bit, $l^{(i)}$, corresponding to the output $y$
and the channel gain $a$, is given by

\begin{equation}
\label{eq2}
l^{(i)}=\log
\frac{p(y|b^i(x)=0,a)}{p(y|b^i(x)=1,a)}=
\log\frac{\sum_{x \in
\chi_0^i}p(y|x,a)}{\sum_{x \in
\chi_1^i}p(y|x,a)} \stackrel{\Delta}{=} g_a^i(y)\:,
\end{equation}
where $b^i(x), i\in \{1,\ldots,m\}$, is the $i$th bit of
the signal $x$, $\chi_w^i$ is the subset of the signals $x$ in
$\chi$ where $b^i(x)=w$, $w\in\{0,1\}$, and the conditional pdfs are given by
$p(y|x,a)=\frac{1}{2\pi\sigma^2}\exp(-\frac{||y-ax||^2}{2\sigma^2})$,
for two-dimensional signal constellations, or by
$\frac{1}{\sqrt{2\pi}\sigma}\exp(-\frac{(y-ax)^2}{2\sigma^2})$
for one-dimensional constellations.

For the case of BPSK modulation, the above formulation simplifies to:
\begin{equation}
\label{eq3}
l =
%\log \frac{f_{Y|X,A}(y|x=+1,a)}{f_{Y|X,A}(y|x=-1,a)}=
\frac{2a}{\sigma^2}y\:.
\end{equation}
%where we have used
%\begin{equation}
%\label{eq22}
%f_{Y|X,A}(y|x,a)=\frac{1}{\sqrt{2\pi}\sigma}\:
%\exp(-\frac{(y-ax)^2}{2\sigma^2})\:.
%\end{equation}

\item \emph{Unknown CSI}: In this case, the
channel gain $A_t$ is unavailable at the receiver. The channel LLR for the $i$th bit can
then be calculated as

\begin{equation}
\label{eq4}
l^{(i)}=\log
\frac{p(y|b^i(x)=0)}{p(y|b^i(x)=1)}=
\log\frac{\sum_{x\in
\chi_0^i}p(y|x)}{\sum_{x \in
\chi_1^i}p(y|x)} \stackrel{\Delta}{=} g^i(y),
\end{equation}
where
$p(y|x)=\int_{0}^{\infty}\frac{1}{2\pi\sigma^2}\exp(-\frac{||y-ax||^2}{2\sigma^2})\,p_A(a)da$,
for two-dimensional constellations, or $p(y|x)=\int_{0}^{\infty}\frac{1}{\sqrt{2\pi}\sigma}\:
\exp(-\frac{(y-ax)^2}{2\sigma^2}) p_A(a)da$, for one-dimensional constellations
(including BPSK).
%and $g^i(\mathbf{y})$ represents $L^i$ as a function of
%$\mathbf{y}$.

%For the case of BPSK, the above formulation simplifies to:
%
%\begin{equation}
%\label{eq5}
% l= \log \frac{\int_{0}^{\infty}f_{Y|X,A}(y|x=+1,a)\:f_A(a)\;da}{\int_{0}^{\infty}f_{Y|X,A}(y|x=-1,a)\:f_A(a)\;da}\:,
%\end{equation}
%where $f_{Y|X,A}(y|x,a)$ is given in (\ref{eq22}).

\begin{itemize}
\item {\em BPSK}

For BPSK modulation over normalized Rayleigh fading channels,
Equation (\ref{eq4}) reduces to~\cite{YA-09}
\begin{equation}
\label{eq6}
l=\log \frac{\Phi(y/\sqrt{2\sigma^2(1+2\sigma^2)})}{\Phi(-y/\sqrt{2\sigma^2(1+2\sigma^2)})},
\end{equation}
where $\Phi(z)=1+\sqrt{\pi}ze^{z^2}\mathrm{erfc}(-z)$, and
$\mathrm{erfc}(.)$ represents the complementary error function.

\item {\em M-ary PAM}

For the $M$-ary PAM signal set, the conditional pdf of the received signal, assuming a normalized Rayleigh fading channel with no CSI
is given by
%\begin{displaymath}
%p(y|x)=\int_{0}^{\infty}\frac{1}{\sqrt{2\pi}\sigma}\exp(-\frac{(y-ax)^2}{2\sigma^2})\,p_A(a)da\:,
%\end{displaymath}
%which reduces to
\begin{align}
\label{eq8}
p(y|x)=\frac{e^{-(y^2/\hat{\sigma}^2)}}{\sqrt{\pi}\hat{\sigma}^{3}}&(\sqrt{\pi}xy\,
\mathrm{erfc}(-\frac{\sqrt{2}xy}{2\sigma\hat{\sigma}})+ \notag
\\&\sqrt{2}\sigma \hat{\sigma}\,e^{-\frac{x^2y^2}{2\sigma^2
\hat{\sigma}^2}}),
\end{align}
where $\hat{\sigma}=\sqrt{x^2+2\sigma^2}$. Replacing  (\ref{eq8}) in~(\ref{eq4}), the true LLR values can be calculated.

\item {\em $M$-ary QAM}

For the $M$-ary QAM constellation, assuming the normalized Rayleigh fading channel with no CSI at
the receiver, we have the following conditional pdf of the received complex value $y=y_{r}+jy_{i}$ given
that the symbol $x=x_{r}+jx_{i}$ is transmitted:
\begin{align}
\label{eq::qam}
p(y|x)
%&=\int_{0}^{\infty}\frac{1}{2\pi\sigma^2}\exp(-\frac{||y-ax||^2}{2\sigma^2})p_A(a)da\notag \\
&=\bigg(\frac{1}{\sqrt{2\pi}}
(y_{r}x_{r}+y_{i}x_{i})e^{\frac{x_{r}x_{i}y_{r}y_{i}}{\sigma^2\hat{\sigma}^2}}\mathrm{erfc}(-\frac{\sqrt{2}(x_{r}y_{r}+x_{i}y_{i})}{2\sigma\hat{\sigma}})\notag\\
&+\frac{\sigma\hat{\sigma}}{\pi}e^{-\frac{(x_r^2y_r^2+x_i^2y_i^2)}{2\sigma^2\hat{\sigma}^2}}\bigg)\frac{1}{\sigma\hat{\sigma}^3}e^{-\frac{\gamma^2}{2\sigma^2\hat{\sigma}^2}},
\end{align}
where $\gamma^2=x_r^2y_i^2+x_i^2y_r^2+2\sigma^2(y_r^2+y_i^2)$ and
$\hat{\sigma}=\sqrt{x_r^2+x_i^2+2\sigma^2}$.

True LLR values are then calculated using (\ref{eq4}) and (\ref{eq::qam}).
\end{itemize}

\end{enumerate}

In this paper, our main focus is on Case $2$, where the CSI is unknown
at the receiver. In this case, the relationship between the channel
LLR(s) and the channel output, given by (\ref{eq4}), is rather complex.
This means that the calculation of channel LLR values, required at
the receiver, as a function of channel outputs is computationally
expensive. Moreover, it would be very challenging to obtain the pdf
of the channel LLR using (\ref{eq4}). This pdf would be helpful to
analyze the performance of different detection/decoding algorithms
or to design one. In~\cite{hou}, the linear approximation of
(\ref{eqsa}) was proposed to simplify the relationship between $L$
and $Y$ for the BPSK modulation. This approximation however has proved to be rather
inaccurate resulting in performance degradations of a few tenths of
a dB compared to true LLR calculations~\cite{YA-09}. Recently, the
more accurate linear approximation of (\ref{eqtl}) was proposed
in~\cite{YA-09}. This approximation was shown in~\cite{YA-09} to
perform well with binary LDPC codes and the BPSK modulation, and to result in rather large
performance improvements compared to the approximation of
(\ref{eqsa}). The downside however is the complicated relationship
between $\alpha_{\hat{C}}$ and $\sigma^2$, and the relatively high
computational complexity of solving (\ref{eqoj}).
The approach of~\cite{YA-09} was then generalized in~\cite{YA-10}
to non-binary constellations, where piece-wise linear approximations
of LLR were devised. These approximations performed very closely to true
LLR calculations for an LDPC-coded BICM scheme based on the $8$-PAM constellation,
but relatively poorly for the $16$-QAM constellation (a gap of about 0.2 dB
to the true LLR calculations).

It is important to note that other approximations of the channel LLR for fading
channels are given in the literature. For example, in~\cite{TB-02}, the {\em log-sum
approximation} $\log \sum_k \beta_k \approx \max_k \log \beta_k$ was used
to approximate (\ref{eq2}) as a piece-wise linear function. The approximation however
is only good in the high SNR regime, where the sum is dominated by a single
large term. Moreover, in the absence of CSI, which is of interest in this paper,
the approximation does not lead to a piece-wise linear function and is much
more complicated to implement.

Also noteworthy is that the LLR approximation (\ref{eqsa}) of~\cite{hou} for BPSK with unknown CSI
can be interpreted as the minimum mean square error (MMSE) estimation of (\ref{eq3}), 
the BPSK LLR for known CSI, given the received value $y$. The idea of using the MMSE estimate of LLR 
for known CSI as the approximate LLR in the absence of CSI can also be applied to non-binary modulations. 
Our study however shows that the performance degradation compared to true LLR calculations is 
rather large, e.g., about $0.5$ dB for both $8$-PAM and $16$-QAM. 

\subsection{LDPC Coding}
We consider the application of binary LDPC codes for the transmission of information
over the binary-input uncorrelated flat Rayleigh fading channel described
in Subsection~\ref{subsec2.1}. This channel is {\em memoryless}.
It is also {\em output-symmetric}~\cite{rich01B} when BPSK modulation is used for the transmission.
It is however known that when $M > 2$, the $m$ bit-channels associated with
the BICM scheme are not necessarily output-symmetric~\cite{HSMP-03}. To simplify the analysis,
we thus use the technique of augmenting the bit-channels with i.i.d.
channel adapters as in~\cite{HSMP-03}. This makes the resulting channels output-symmetric.
We also consider {\em symmetric message-passing decoders}~\cite{rich01B} such that the conditional
error probability is independent of the transmitted codeword~\cite{rich01B}.
For simplicity, therefore, we assume that the all-zero codeword is transmitted.
This will be particularly helpful for density evolution, where the pdf of
LLR values and their approximations is needed under this assumption.

\begin{figure}[h]
\centering
\includegraphics[width = 0.7\textwidth]{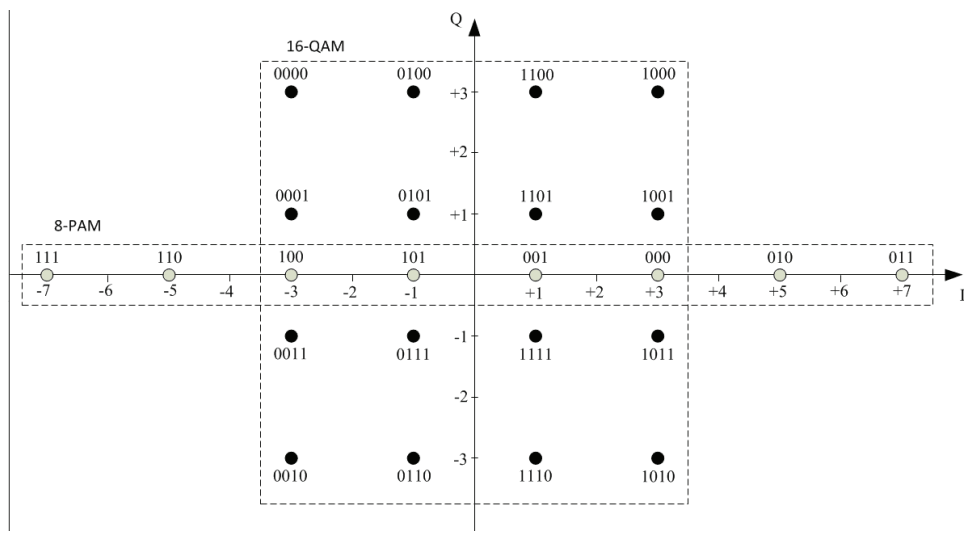}
\caption{$8$-PAM and $16$-QAM constellations with Gray
labeling} \label{fig::const}
\end{figure}

The error rate performance of the transmission schemes is measured as a function of
the channel signal-to-noise ratio (SNR), given by $E_b/N_0$ for BPSK,
and $E_s/N_0$ for non-binary modulations, where $E_b$ and $E_s$ are the
average energy per information bit and per transmitted symbol, respectively,
and $N_0$ is the one-sided power spectral density of the additive white
Gaussian noise (AWGN). For BPSK, assuming $X= \pm 1$ is transmitted, the SNR is related to the
noise variance by $E_b/N_0 = 1/(2R\sigma^2)$, where $R$ is the rate of the LDPC code.
For $8$-PAM and $16$-QAM, given in Fig.~\ref{fig::const}, the SNR is given
by $21/(2 \sigma^2)$, and  $10/(2 \sigma^2)$, respectively.

To investigate the performance of transmission schemes, we use Monte-Carlo simulations
at finite block lengths, and density evolution for asymptotic analysis. In the latter case,
the threshold~\cite{rich01B},~\cite{hou},~\cite{HSMP-03} of the transmission scheme for LDPC code ensembles is
calculated as the measure of performance. For more information
on LDPC codes and the calculation of the threshold, the reader is referred
to~\cite{rich01B},~\cite{hou},~\cite{HSMP-03}.

\section{LLR APPROXIMATION BASED ON TAYLOR SERIES}
\label{sec::llr}
\subsection{Brief review of Taylor series}
%In accordance with the LLR calculation of non-binary constellations
%5the true LLRs are complicated nonlinear functions. Since these
%functions have $n$ continuous derivatives in $\mathbb{C}$ field
%therefore the Taylor polynomials converge to the original function.
In this subsection, we provide the definition of one- and two-dimensional Taylor series (polynomials),
which we subsequently use for the LLR approximation of one-dimensional and
two-dimensional signal constellations, respectively.
%from\cite{DD-10,GL-10}.
\begin{definition}
Suppose that $f(x)$ has $n$ derivatives at a point $x_0 \in [a,b]$. The
one-dimensional Taylor polynomial of order $n$ for $f(x)$ at point
$x_0$ is then defined by
\begin{align}
P_n(x)&=f(x_0)+f^{\prime}(x_0)(x-x_0)+\frac{f^{\prime\prime}(x_0)}{2}(x-x_0)^2+\ldots+\frac{f^{(n)}(x_0)}{n!}(x-x_0)^n
\notag \\
&=\sum_{k=0}^{n}\frac{f^{(k)}(x_0)}{k!}(x-x_0)^k.
\end{align}
\end{definition}

\begin{definition}
Suppose that $f(x,y)$ has up to $n$th partial derivatives at a point $(x_0,y_0)$ on a
convex subset $\Omega$ of $\mathbb{R}^2$. The two-dimensional Taylor polynomial of
order $n$ for $f(x,y)$ at $(x_0,y_0)$ is then defined by
\begin{equation}
 \label{eq::biTaylor}
 P_n(x,y)=\underset{\ell+m\leq n}{\sum_{\ell\geq 0}\sum_{m\geq
 0}}\frac{\partial^{m+\ell}f}{\partial x^{\ell} \partial
 y^m}(x_0,y_0)\,\frac{(x-x_0)^{\ell}}{\ell!}\frac{(y-y_0)^{m}}{m!}\quad
 \text{{\em for any}}\; (x,y)\in \Omega.
 \end{equation}
\end{definition}

The difference between the true value of the function and its Taylor polynomial approximation
is called the {\em remainder}. There are different forms to represent the remainder. The most commonly
used is the {\em Lagrange form}, where the reminder in the case of one- and two-dimensional
Taylor series, is given by~\cite{B-76}
\begin{equation}
\frac{f^{(n+1)}(c)}{(n+1)!}(x-x_0)^{n+1}\:,
\label{eqdu}
\end{equation}
and
\begin{equation}
\underset{\ell+m = n+1}{\sum_{\ell\geq 0}\sum_{m\geq
 0}}\frac{\partial^{m+\ell}f}{\partial x^{\ell} \partial
 y^m}(x_1,y_1)\,\frac{(x-x_0)^{\ell}}{\ell!}\frac{(y-y_0)^{m}}{m!}\:,
\label{eqyv}
\end{equation}
respectively, where $c$ is a number between $x_0$ and $x$, and
$(x_1,y_1)$ is a point on the line connecting $(x_0,y_0)$ and
$(x,y)$. For the one-dimensional case, this requires $f^{(1)},
f^{(2)}, \ldots, f^{(n)}$ to be continuous on $[a,b]$,
and that $f^{(n+1)}$ to exist in $(a,b)$~\cite{B-76}.
For the two-dimensional case, to have (\ref{eqyv}) as the remainder, the requirement is that $f$
must have continuous partial derivatives of order $n+1$ in a neighborhood
of every point on a line segment joining two points $(x_0,y_0)$ and
$(x,y)$ in $\Omega$~\cite{B-76}. More details on the convergence properties of
Taylor series can be found in~\cite{B-76},~\cite{DD-10},~\cite{GL-10}.

\subsection{LLR approximations}

Our goal in this part of the paper is to approximate the LLR for one- and two-dimensional signal constellations
as a function of the channel output, using Taylor polynomials of different orders. For the channel model considered here,
the LLR is a differentiable function of the channel output and thus lends itself well to the Taylor
series approximation. We are interested in using polynomials of smallest order as long as the approximation is accurate enough
for the application under consideration. In particular, linear approximations
are of most interest followed by second, third and larger order approximations. To obtain a sufficiently accurate approximation over a wide range of channel outputs
and for a relatively low complexity, it is sometimes beneficial to use piece-wise approximations,
where the domain of interest is partitioned into sub-domains. In each sub-domain then, a different
approximation will be used. While it is possible to approximate the LLR within a given domain of interest
with an arbitrarily high accuracy, if one has no constraints on the number of
sub-domains and the order of approximations; in practice, due to the limited computational resources,
the number of sub-domains and the order of approximations need to be small.

In addition to the selection of the number and the boundaries of the sub-domains, and the order of approximation
in each sub-domain, one also needs to choose the point within each sub-domain at which the
Taylor series is derived. This in general, could be a complicated optimization problem,
defeating the whole purpose of finding low-complexity approximations for LLR. In this work,
we limit ourselves to (piece-wise) linear, second and third order approximations. Since the accuracy of the LLR
is particularly important for smaller LLR values, where a small error can change the sign of the LLR and thus
the corresponding hard-decision, we select the roots of the LLR as the points around which the Taylor series
is derived. The number and boundaries of the sub-domains will then be identified based on the number of roots,
and the general shape of the LLR function. This will be explained in more details for BPSK, $8$-PAM
and $16$-QAM in the following.\footnote{BPSK, $8$-PAM and $16$-PSK are all used in~\cite{YA-09}, and~\cite{YA-10}. We thus use
the same constellations for comparison.} In particular, the simplicity of BPSK modulation makes it possible to
derive closed-form analytical expressions for the Taylor approximations for any value of the channel SNR.

\subsubsection{BPSK Modulation}
The uncorrelated flat Rayleigh fading channel with BPSK modulation is output-symmetric.
For an output-symmetric channel, LLR is an odd function of the
channel output~\cite{YA-09}, i.e., if $L = g(Y)$, then $g(-y) =
-g(y),\:\forall y$. It is thus natural to look for approximations
which maintain the same odd symmetry. Both linear approximations
(\ref{eqsa}) and (\ref{eqtl}) are odd symmetric. From (\ref{eq6}),
it is easy to see that for the channel under consideration, the LLR
is a continuous and differentiable function of the channel output
with only a single root at $y=0$. In fact, the Taylor series of (\ref{eq6}) in the neighborhood of
$y=0$ is given by
\begin{eqnarray}
\label{eq11}
\sqrt{\frac{2\pi}{1+2\sigma^2}}\:\frac{y}{\sigma}+\frac{\sqrt{2\pi}(\pi-3)}{6(1+2\sigma^2)^{3/2}}(\frac{y}{\sigma})^3 + O(y^5)\:.
\end{eqnarray}
Corresponding to (\ref{eq11}), we propose the following linear approximation of
the channel LLR:
\begin{equation}
\label{eq12}
\hat{L}_{LT} = \sqrt{\frac{2\pi}{1+2\sigma^2}}\:\frac{Y}{\sigma} \stackrel{\Delta}{=}
\alpha_{T} Y\:,
\end{equation}
and the non-linear approximation given by
\begin{equation}
\label{eqrd}
\hat{L}_{NT} = \sqrt{\frac{2\pi}{1+2\sigma^2}}\:\frac{Y}{\sigma} +
\frac{\sqrt{2\pi}(\pi-3)}{6(1+2\sigma^2)^{3/2}}(\frac{Y}{\sigma})^3 \stackrel{\Delta}{=}
\alpha_{T} Y + \beta_{T} Y^3\:.
\end{equation}
Note that for small channel SNR values where $2\sigma^2 >> 1$,
the approximation (\ref{eq12}) reduces to (\ref{eqsa}).

\begin{figure}[h]
\centering
\includegraphics[width = 0.7\textwidth]{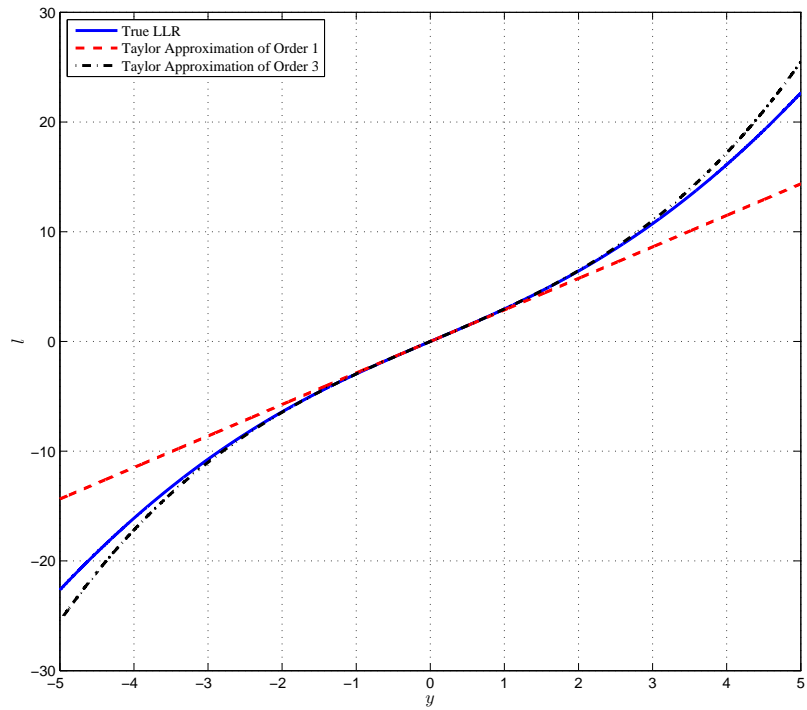}
\caption{Comparison of true LLR values and the approximations obtained by Taylor series
for the uncorrelated flat Rayleigh fading channel with unknown CSI and $\sigma=0.6449$.}
\label{figvb}
\end{figure}

In Fig.~\ref{figvb}, true LLR values from (\ref{eq6}) are compared with the proposed approximations
(\ref{eq12}) and (\ref{eqrd}) for $\sigma = 0.6449$. As expected, the approximations are very accurate for
smaller values of $y$, with the non-linear approximation being more accurate
and almost identical to the true LLR values for $|y| <4$.

To apply density evolution to iterative algorithms with the approximated channel LLR values, we need to derive the pdf of $\hat{L}_{LT}$
and $\hat{L}_{NT}$ given in (\ref{eq12}) and (\ref{eqrd}), respectively, assuming
$X = +1$ is transmitted. From (\ref{eq2}), we have
\begin{equation}
\label{eq14}
p(y|+1,a)=\frac{1}{\sqrt{2\pi}\sigma}\exp(-\frac{(y-a)^2}{2\;\sigma^2})\:.
\end{equation}
Averaging (\ref{eq14}) over the distribution of $A$, we obtain
\begin{equation}
\label{eq15}
p(y|+1)=\frac{\sqrt{2/\pi}\:\sigma}{1+2\sigma^2}\exp({-\frac{y^2}{1+2\sigma^2}})\Theta(\frac{y}{
\sqrt{2\sigma^2(1+2\sigma^2)}}),
\end{equation}
where $\Theta(z)=\:e^{-z^2}+\sqrt{\pi}z\:\text{erfc}(-z)$. Based on (\ref{eq12}),
the pdf of $\hat{L}_{LT}$, given that $X = +1$ is transmitted, is derived as:
\begin{eqnarray}
\label{eq13}
p_{\hat{L}_{LT}{\{\sigma\}}}(\hat{l})=\frac{\sigma^2}{\pi\:\sqrt{1+2\sigma^2}}\bigg(\exp{(-\frac{(1+2\sigma^2)\hat{l}^2}{4\pi})}
+ \\ \nonumber
\frac{1}{2} \hat{l}\:\exp{(-\frac{\sigma^2\;\hat{l}^2}{2\pi})}\:\text{erfc}(-\frac{\hat{l}}{2\sqrt{\pi}})\bigg).
\end{eqnarray}

The pdf of $\hat{L}_{NT}$ is calculated using (\ref{eq15}), and based
on the relationship (\ref{eqrd}) between $\hat{L}_{NT}$ and $Y$ as follows:
\begin{equation}
\label{eq16}
p_{\hat{L}_{NT}{\{\sigma\}}}(\hat{l})=
\frac{p(y|+1)\Big|_{y=g^{-1}(\hat{l})}}{|g^\prime(y)|\Big|_{y=g^{-1}(\hat{l})}},
\end{equation}
where $g(y) = \alpha_{T} y + \beta_{T} y^3$ with the derivative
$g^\prime(y) = \alpha_{T} + 3 \beta_{T} y^2$, and the inverse
\begin{equation}
\label{eq17}
 g^{-1}(\hat{l})= \frac{\phi(\hat{l})}{6\;\beta_T}-\frac{2\alpha_T}{\phi(\hat{l})},
\end{equation}
in which $\phi(\hat{l})=\sqrt[3]{12\:\beta_T^2(9\hat{l}+\sqrt{\frac{12\:\alpha_T^3}{\beta_T}+81\:\hat{l}^2})}$.

\subsubsection{$8$-PAM}
\label{subsecfd}
\begin{figure}[h]
\centering
\subfloat[$i=1$]{\label{fig3:bit1}\includegraphics[width = 0.4\textwidth]{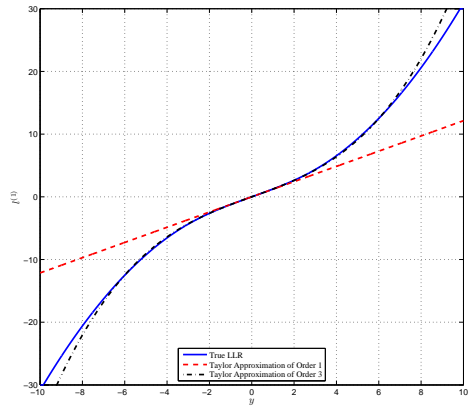}}\\
 \centering
\subfloat[$i=2$]{\label{fig3:bit2}\includegraphics[width = 0.4\textwidth]{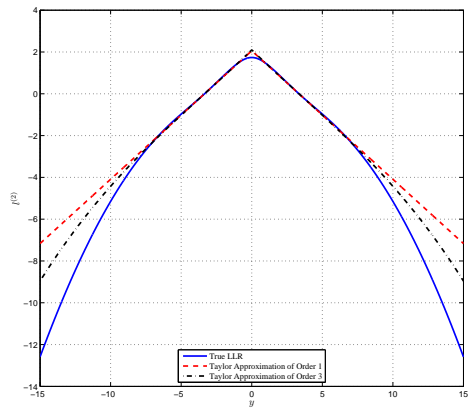}}\\
\centering
\subfloat[$i=3$]{\label{fig3:bit3}\includegraphics[width = 0.4\textwidth]{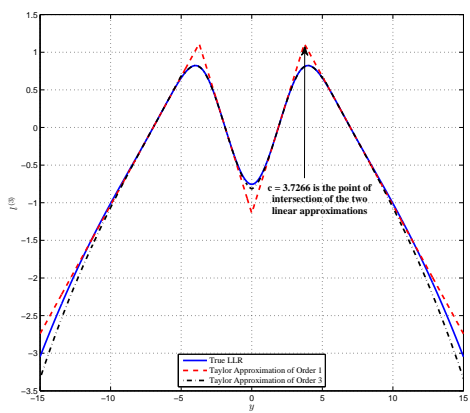}}\\
 \caption{True bit LLR values $l^{(1)}, l^{(2)}$, and $l^{(3)}$ as functions of the channel output $y$ for $8$-PAM at SNR=$7.91$ dB,
along with the corresponding (piece-wise) Taylor approximations.}
\label{fig3}
\end{figure}

Consider the $8$-PAM signal set with Gray labeling as shown in Fig.~\ref{fig::const}.
True LLR values for the three bits can be calculated using (\ref{eq4}) and (\ref{eq8}).
These values clearly depend on the value of the channel SNR. We now explain the
Taylor approximation of the three LLRs based on a given value of SNR$=7.91$ dB.
In Fig.~\ref{fig3}, for SNR$=7.91$ dB, the true LLR values are plotted with full lines as a function of the channel output $y$.
As expected, all three functions are symmetric with respect to the vertical axis at $y=0$.
The LLR of the first bit, $l^{(1)}$, has a single root at $y=0$. The roots of $l^{(2)}$ and $l^{(3)}$
are respectively $\{-3.3449, 3.3449\}$ and $\{-6.9832, -1.8848, 1.8848, 6.9832\}$.
We thus partition the domain of $y$ values to one, two and four sub-domains, for the three LLRs,
respectively.

Suppose that the $k$th derivative of the LLR function of the $i$th bit is denoted
by $f^{(k)}_i$. For the first bit, a linear approximation of the LLR can be obtained by the first order Taylor
polynomial at $y=0$ as follows:
\begin{equation}
\hat{L}^{(1)} = f^{(1)}_1(0) Y\:.
\label{eqwh}
\end{equation}
Due to the symmetry of the LLR functions, we have
\begin{align}
f^{(2k-1)}_{i}(y) &= -f^{(2k-1)}_{i}(-y),\notag \\
f^{(2k)}_{i}(y) &= f^{(2k)}_{i}(-y)\:,
\end{align}
for integers $k \geq 1$.
We thus have the following piece-wise linear approximations based on Taylor series for the
second and the third bits, respectively:
\begin{equation}
\hat{L}^{(2)} = f^{(1)}_2(y_1)(Y-y_1) I(Y\geq0) - f^{(1)}_2(y_1)(Y+y_1) I(Y\leq0) = f^{(1)}_2(y_1)|Y|-f^{(1)}_2(y_1)y_1\:,
\label{eqrj}
\end{equation}
\begin{align}
\hat{L}^{(3)}&=f^{(1)}_3(y_2)(Y-y_2) I(c\leq Y)+f^{(1)}_3(y_3)(Y-y_3) I(0\leq Y<c)\notag\\
&-f^{(1)}_3(y_3)(Y+y_3) I(-c < Y\leq0)-f^{(1)}_3(y_2)(Y+y_2) I(Y\leq -c)\notag\\
&=\left(f^{(1)}_3(y_2)|Y|-f^{(1)}_3(y_2)y_2\right) I(c\leq |Y|)+\left(f^{(1)}_3(y_3)|Y|-f^{(1)}_3(y_3)y_3\right) I(|Y|<c)\:.
\label{eqrh}
\end{align}
where $I(.)$ is the indicator function, $y_1 = 3.3449$, $y_2 =  6.9832$, $y_3 = 1.8848$,
%$f^{(1)}_i(y_{z_k})$, $1\leq i \leq 3$ denotes the first derivative
%of LLR function of $i$-th bit in $y_{z_k}$,
and $c = 3.7266$ is the $y$ value of the point where the two linear approximations for the LLR
function of the third bit intersect in the $y > 0$ region (see Fig. \ref{fig3:bit3}).

In Table~\ref{tabrd}, the coefficients of the linear approximation~(\ref{eqwh}) and the piece-wise linear approximations~(\ref{eqrj}) and (\ref{eqrh})
for SNR$=7.91$ dB are given in the first row. It is worth mentioning that for all three functions,
second order derivatives at the roots of LLR functions are zero. The coefficients of the terms with degree $3$ in the
Taylor polynomial are also given in the second row of the table.

In Fig.~\ref{fig3}, the first, and the third order Taylor polynomials for the approximation of
the three LLRs are also plotted. Comparison with true LLRs shows that the accuracy of the
approximations improve consistently by increasing the order of the polynomials.

\subsubsection{$16$-QAM}
\label{subsecrf}
Consider the $16$-QAM constellation with Gray labeling as shown in Fig.~\ref{fig::const}.
True LLR values for the four bits can be calculated using (\ref{eq4}) and (\ref{eq::qam}).
These values clearly depend on the value of the channel SNR. We now explain the
Taylor approximation of the four LLRs based on the value of SNR$=4.89$ dB.

Since $16$-QAM is two-dimensional, we need to use two-dimensional Taylor polynomials for the approximation of LLRs,
where each LLR is a function of the channel output $y=y_r + j y_i$.
One difference compared to the one-dimensional cases is that, the roots of the
LLR functions in the two-dimensional case are located on one or more two-dimensional curves rather than
belonging to a discrete set of values. To derive the Taylor polynomial in a sub-domain, a single point from
such a curve within the sub-domain should then be selected so that the Taylor coefficients can be
computed at that point. The selection of such a point depends on the general shape and symmetries of the LLR function, as
explained in the following.

In Figures~\ref{fig5}(a) and (b), the contours of fixed (true) LLR values for the first and the second bit
of the constellation are shown in the $(y_r,y_i)$ plane, respectively.
As can be seen, for the first bit, the curve corresponding to $l^{(1)} = 0$ is $y_r = 0$.
For the second bit, there are two curves, symmetric with respect to $y_r = 0$ corresponding to $l^{(2)} = 0$.
The existing symmetries in the LLR functions with respect to both $y_r = 0$ and $y_i = 0$ suggest the selection
of $(0,0)$, and $(\xi,0)$ and $(-\xi,0)$, as the points around which the Taylor approximations
of $l^{(1)}$ and $l^{(2)}$, should be derived, respectively, where $\xi = 1.8908$ is the intersection
of the curve corresponding to $l^{(2)} = 0$ with the line $y_i = 0$ in the region $y_r > 0$ of the $(y_r,y_i)$ plane.

The corresponding Taylor polynomials of the third and the second order for
$L^{(1)}$ and $L^{(2)}$, are respectively derived as:
\begin{align}
\hat{L}^{(1)}&=
-0.9878\,Y_r -0.04285\,Y_r\,Y_i^2-0.01654\,Y_r^3\:,
\label{eqgv}
\end{align}
\begin{align}
\hat{L}^{(2)}=&
-0.9285+0.2690\,|Y_r|+0.1174\,Y_r^2-0.0364\,Y_i^2\:.
\label{eqpd}
\end{align}
Based on the symmetry in the Gray labeling of $16$-QAM, the LLR values for the third and the fourth bits
are similar to those of the first and the second bits, respectively, except that the real and the imaginary parts of
$y$ need to be switched. We thus have the following Taylor approximations for $L^{(3)}$ and $L^{(4)}$, respectively:
\begin{align}
\hat{L}^{(3)}&=
0.9878\,Y_i+0.04285\,Y_i\,Y_r^2+0.01654\,Y_i^3\:,
\label{eq::biTbit13}
\end{align}
\begin{align}
\hat{L}^{(4)}=&
-0.9285+0.2690\,|Y_i|+0.1174\,Y_i^2-0.0364\,Y_r^2\:.
\label{eq::biTbit24}
\end{align}

\section{NUMERICAL RESULTS AND DISCUSSION}
\label{sec::simul} In this paper, we mainly compare our results with
those of~\cite{YA-09} and~\cite{YA-10}, which are the best known approximations of the channel LLR for uncorrelated flat Rayleigh fading
channels in terms of performance. Similar to~\cite{YA-09} and~\cite{YA-10}, we consider the three modulations BPSK, 8-PAM and 16-QAM.
Our results however, can be easily extended to other linear modulations.

To obtain the coefficients of Taylor approximations, we resort to the asymptotic analysis of density evolution with the SNR threshold as the performance criterion.
The goal is then to find sufficiently accurate Taylor approximations for the LLR functions so that the
resulting threshold of the BICM scheme is close to the threshold obtained using the true LLR
values. As the Taylor coefficients are functions of the channel SNR, one approach would be to
start from an SNR value above the threshold, and find the corresponding Taylor coefficients. Then
find the threshold corresponding to the resulting Taylor approximation. (This threshold will be
smaller than the starting SNR.) Use the new SNR threshold and find the corresponding Taylor approximation.
Use this new approximation to find the next threshold. Continue this process until it converges to a
fixed point, i.e., the Taylor coefficients and the SNR threshold remain unchanged in two successive iterations.
A simpler approach, with slightly inferior results, is to find the SNR threshold using the true
LLR values and then use that value of SNR to find the Taylor coefficients.

\subsection{BPSK}
In this part, we present analysis and design results based on the proposed linear and non-linear
LLR approximations for the BPSK modulation through a number of examples.

\begin{example}
\label{ex2} In this example, decoding thresholds\footnote{The thresholds are given in terms of $E_b/N_0$.} of two regular
ensembles of LDPC codes based on different LLR approximations over
the uncorrelated flat Rayleigh fading channel are calculated. The
ensembles have the following degree distributions:
$\lambda_1(x)=x^2,\:\rho_1(x)=x^5$;
$\lambda_2(x)=x^3,\:\rho_2(x)=x^{15}$.\footnote{The threshold
results presented in this paper are obtained for an 11-bit iterative
belief propagation decoder~\cite{rich01B} with the maximum number of
iterations 1000. The target probability of error for variable node
to check node message error rate is set at $10^{-7}$. The maximum
LLR is chosen to be $25$ except for simulations related to
non-linear Taylor approximation, for which the maximum LLR of $35$
is used.} The results corresponding to the two ensembles are given
in Tables \ref{table1} and \ref{table2}, respectively. These results
indicate that while there is a large performance gap between the
linear approximations (\ref{eqsa}) and (\ref{eqtl}), the performance
of the proposed linear approximation (\ref{eq12}) is practically
identical to that of (\ref{eqtl}). Unlike (\ref{eq12}), however, the
computational complexity of (\ref{eqtl}) is relatively high. Also
noteworthy is the fact that the proposed approximation is analytical
while the approximation (\ref{eqtl}) of \cite{YA-09} must be
obtained numerically. In both tables, the results of optimal linear
approximation ($\hat{L}_{opt} = \alpha_{opt}
Y$),\footnote{$\alpha_{opt}$ is obtained by exhaustive search using
density evolution.} and the non-linear approximation of (\ref{eqrd})
are also given. The results show that both linear approximations
perform practically optimally for these codes. Our proposed
non-linear approximation performs the same as the optimal linear
approximation and does not provide any further improvement.
\end{example}
\begin{example}
\label{ex3} In this example, decoding thresholds for two irregular
LDPC code ensembles are calculated. The first ensemble has rate
$1/2$ and is optimized for a normalized Rayleigh fading channel with
known CSI (first code in Table I of \cite{hou}). Using
approximations (\ref{eqsa}) and (\ref{eqtl}), the $E_b/N_0$
threshold for unknown CSI is $3.74$ dB and $2.98$ dB, respectively.
For the proposed linear and non-linear approximations, the
thresholds are $2.98$ dB and $2.97$ dB, respectively.

The second ensemble is a rate-$1/2$ threshold optimized ensemble for
normalized Rayleigh fading channel with unknown CSI with
approximation (\ref{eqtl}) (Code 2 of \cite{YA-09}). Again, for this
ensemble, both (\ref{eqtl}) and the proposed linear approximation
have the same threshold of $2.76$ dB. The threshold for the proposed
non-linear approximation however is improved to $2.73$ dB.
\end{example}

To compare the error rate performance of the proposed linear
approximation and that of \cite{YA-09} at finite block lengths, we
have tested a number of regular and irregular LDPC codes. For all
cases, the two approximations performed practically the same, and
very close to the performance with true LLR calculations. One such
example can be found in~\cite{RBA-10}.

\begin{example}
\label{ex4} In this example, we construct irregular LDPC code ensembles
optimized for uncorrelated flat Rayleigh fading channels with
unknown CSI based on the proposed non-linear approximation
(\ref{eqrd}). To fairly compare our results with those
of~\cite{YA-09}, we choose the exact same constituent degrees as
those in similar examples of~\cite{YA-09}, given in Table III
of~\cite{YA-09}. We design two ensembles, one over a channel with
$\sigma = 0.7436$ where we maximize the rate, and the other with the
fixed rate of $1/2$ where we optimize the threshold. Similar to the
nomenclature used in~\cite{YA-09}, these ensembles are labeled as
``Code 1'' and ``Code 2,'' respectively. The degree distributions for
the two ensembles are given in Table~\ref{table3}. Code 1 has a rate
of $0.4941$ compared to $0.4937$ of the similar code
in~\cite{YA-09}. The threshold of Code 2 is $2.68$ dB which is
$0.08$ dB better than the similar result of \cite{YA-09}.
\end{example}

\subsection{8-PAM}
Similar to \cite{YA-10}, in this part, we consider a BICM scheme with Gray labeled $8$-PAM signal set, as shown in Fig.
\ref{fig::const}, along with a $(3,4)$-regular LDPC code ($(\lambda(x)=x^2,\rho(x)=x^3)$).
The decoding threshold for this scheme using true LLR values is $7.85$ dB.
Using this value of SNR and by following the general piece-wise linear approximation
described in Subsection~\ref{subsecfd}, we then obtain the required Taylor coefficients.
Using this LLR approximation, the decoding threshold is degraded to $7.92$ dB.
If we update the coefficients of the approximation based on SNR$=7.92$ dB, the new threshold will be $7.91$ dB.
The next set of Taylor coefficients, obtained based on SNR$=7.91$ dB, however, do not change the
threshold. This SNR value ($7.91$ dB) is what we used in Subsection~\ref{subsecfd} to derive the
Taylor approximations for the LLR functions of $8$-PAM. We refer the reader to
Subsection~\ref{subsecfd} for the details of the Taylor approximations.

The decoding threshold of the BICM scheme using the first and the third order Taylor polynomials are
$7.91$ dB and $7.86$ dB, respectively. This can be compared to the threshold obtained using the true LLRs, $7.85$ dB,
and the one obtained in~\cite{YA-10}, $7.88$ dB.

\begin{figure}[h]
\centering
\includegraphics[width = 0.7\textwidth]{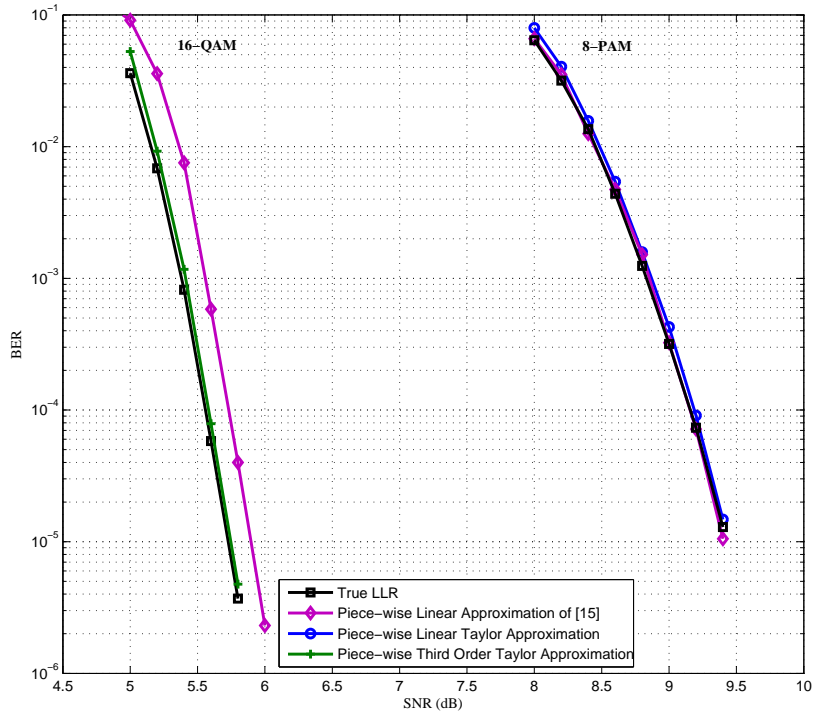}
\caption{BER performances of BICM schemes with $8$-PAM and $16$-QAM in combination with a $(3,4)$-regular LDPC
code of length $12000$ based on various Taylor approximations of LLR, the piece-wise linear approximation of~\cite{YA-10},
and true LLR values, over the uncorrelated flat Rayleigh fading channel.}
\label{fig::FER}
\end{figure}

\begin{figure}[h]
\centering
\subfloat[Contours of fixed $l^{(1)}$]{\label{fig5:bit1}\includegraphics[width = 0.7\textwidth]{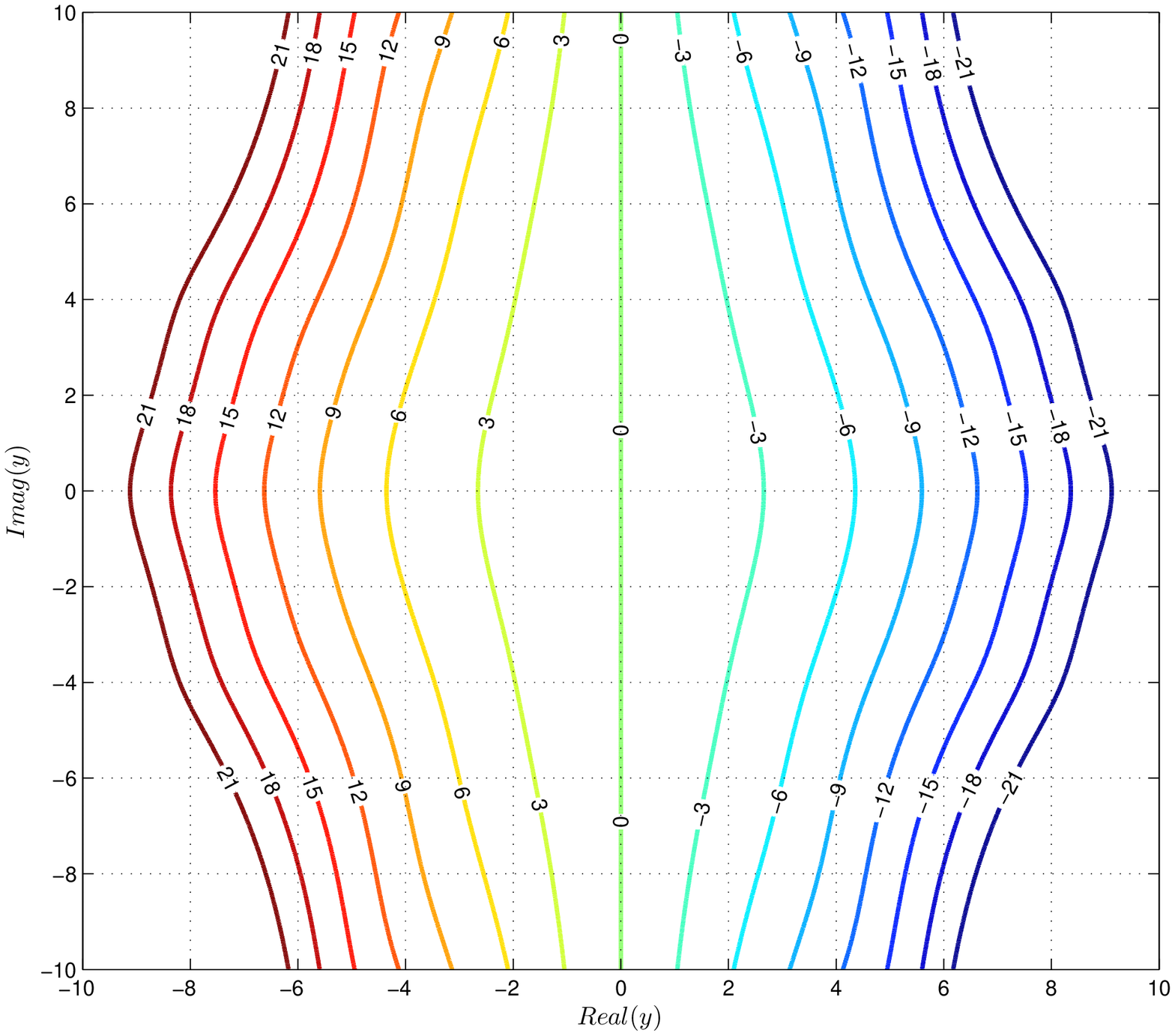}}\\
 \centering
\subfloat[Contours of fixed $l^{(2)}$]{\label{fig5:bit2}\includegraphics[width = 0.7\textwidth]{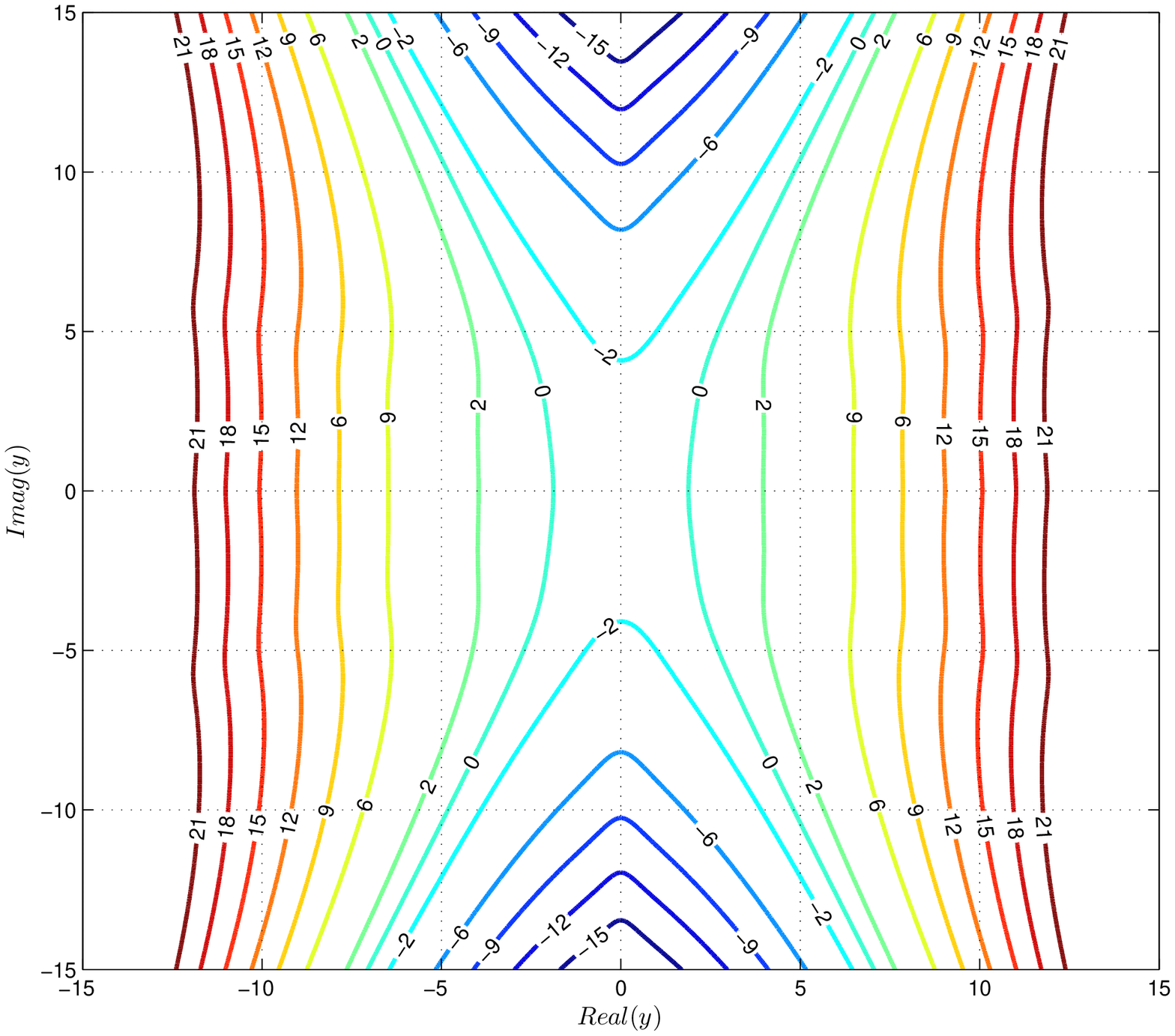}}\\
\caption{Contours of fixed true bit LLR values $l^{(1)}$ and $l^{(2)}$ in the $(y_r,y_i)$ plane for the $16$-QAM at SNR=$4.89$
dB.}
\label{fig5}
\end{figure}

To evaluate the finite length performance of the BICM scheme, we randomly construct a $(3,4)$-regular LDPC code of length 12,000
and girth 6. The bit error rate (BER) performance of this code with the $8$-PAM constellation over the uncorrelated flat Rayleigh fading
channel is shown in Fig. \ref{fig::FER} for both the first and the third order Taylor approximations.
Belief propagation is used for the decoding of the LDPC code with the maximum number
of iterations $100$. In Fig. \ref{fig::FER}, we have also included the BER performance of the same scheme with
true LLR values, and the piece-wise linear approximation of \cite{YA-10} for LLRs.
It is seen that the BER performance of our piece-wise linear approximation is 
similar to the piece-wise linear approximation of~\cite{YA-10}, and almost identical to the 
performance of the more complex true LLRs.
This is while the derivation of our approximation is much simpler than that of \cite{YA-10}.
It can be seen in Fig.~\ref{fig::FER} that no practically significant gain in performance
is obtained by going from first order to the third order Taylor approximation. This however, is not the case
for $16$-QAM, as demonstrated in the next subsection.

\subsection{16-QAM}
In this part, similar to~\cite{YA-10}, we consider the $16$-QAM signal set with Gray labeling
shown in Fig. \ref{fig::const}, and a $(3,4)$-regular LDPC code in the BICM scheme.
To obtain the Taylor approximations, we use the general approach described in Subsection~\ref{subsecrf}.
A Piece-wise linear approximation results in about $0.2$ dB degradation compared to
true LLR values, as also observed in~\cite{YA-10}. We thus consider higher order Taylor polynomials.
The BICM scheme has an SNR threshold of $4.83$ dB with true LLR values.
Based on this SNR value, we then obtain the Taylor coefficients of the third order approximations
for bits one and three and the second order approximations for bits two and four. These approximations are then used
to find the new threshold of the scheme. Repeating this process, we finally converge to
the approximations given in~(\ref{eqgv})-(\ref{eq::biTbit24}) for the four LLRs and the SNR threshold of
$4.89$ dB which is very close to that of true LLRs, $4.83$ dB. If we consider the third order Taylor
polynomial (instead of the second order) for the second and forth bits, the threshold improves slightly
to $4.87$ dB. Note that this provides $0.15$ dB improvement over the $5.02$ dB threshold obtained by the piece-wise linear approximation
of \cite{YA-10} for this scheme. This is while the derivation of the proposed approximation is also simpler than that of \cite{YA-10}.

Finite block length BER performance of the BICM scheme using the Taylor approximations for the LLRs is
shown in Fig. \ref{fig::FER}. The results are for a randomly constructed regular $(3,4)$-LDPC code with length
$12000$ and girth 6. The decoding algorithm is belief propagation with maximum number of iterations $100$.
For comparison, BER curves for the true LLR values, and the piece-wise linear approximation of \cite{YA-10}
are also given in the figure. As can be seen, the proposed Taylor
approximation performs almost the same as true LLRs, and outperforms the approximation of \cite{YA-10}
by about $0.2$ dB. This is while the complexity of computing the Taylor approximations is
much less compared to both the true LLR calculation and the piece-wise linear approximation of \cite{YA-10}.

Although the results reported in this paper are obtained for a
fading channel with known channel SNR at the receiver,
they can also be applied to the case where such information is
unavailable at the receiver. In such a case, if the coding scheme is given
and has a threshold $\sigma^*$ with the proposed Taylor approximation, one would use the
proposed approximate LLR values by substituting $\sigma^*$ in the corresponding Taylor approximation.
On the other hand, if one is interested in the design of a coding scheme,
such as an irregular ensemble of LDPC codes of a given rate, over a channel with
unknown $\sigma$ based on the proposed approximations, one can perform the design, assuming
that $\sigma$ is known, to optimize the threshold. If the threshold
value is $\sigma^*$, one should then use the LLR approximation
by substituting $\sigma$ with $\sigma^*$.

\section{CONCLUSION}
\label{sec::conclusion} In communication systems, the receiver often
requires to calculate the channel LLR for the processing of the
received signal. Over wireless channels, this will have to be
performed almost always at the absence of the full knowledge of CSI.
Under such conditions, the calculation of true LLR values is
computationally expensive. Approximations of LLR, are thus important to find. In this paper, we proposed
simple (piece-wise) linear and non-linear approximations of channel
LLR based on Taylor Series. For the uncorrelated flat Rayleigh
fading channel using one-dimensional linear modulations in the context of LDPC-coded BICM schemes, the proposed (piece-wise) linear approximations perform
practically the same as the best known (piece-wise) linear approximations of~\cite{YA-09} and~\cite{YA-10} and very close to the
performance of a scheme using true LLR values. This is while the derivation of the proposed Taylor approximations for the LLR
is simpler than the computation of the approximations in \cite{YA-09} and~\cite{YA-10}, and significantly simpler than the calculation of true LLR values.
For two-dimensional constellations, where the piece-wise linear approximation of~\cite{YA-10} causes non-negligible
performance loss compared to the true LLR calculations, our proposed Taylor approximations still perform very close
to true LLR calculations with significantly lower complexity.

%Our results on LDPC-coded transmission indicate that
%the proposed approximations can be used in replacement for true LLR values with
%negligible degradation in performance. Similar results are expected to
%hold for other coding and transmission schemes which require the channel LLR
%values for detection and decoding.

\section*{Acknowledgment}
The authors wish to thank Raman Yazdani for providing them with his
ensemble optimization software.

\newpage

\begin{table*}[h]
\center
 \caption{\small{COEFFICIENTS OF TAYLOR POLYNOMIALS FOR THE LLR APPROXIMATIONS OF $8$-PAM}}
\label{table4}
\begin{tabular}[p]{|c|c|c|c|}
\hline
 \centering
  &  Bit $1$ & Bit $2$ & Bit $3$ \\ \hline
  Degree-1 coef. & $f^{(1)}_1(0)=1.2135$  &  $f^{(1)}_2(3.3449)=-0.6147$ & $f^{(1)}_3(6.9832)=-0.3419,f^{(1)}_3(1.8848)=0.6046$ \\  \hline
  Degree-3 coef. & $f^{(3)}_1(0)=0.1420$ & $f^{(3)}_2(3.3449)=0.0039$ & $f^{(3)}_3(6.9832)=-0.0070,f^{(3)}_3(1.8848)=-0.2920$  \\ \hline
  \end{tabular}
\label{tabrd}
\end{table*}

\begin{table*}[h]
\center
 \caption{\small{THRESHOLDS FOR THE LDPC ENSEMBLE WITH $\lambda_1(x)= x^2,\:\rho_1(x)=x^5$ UNDER
BELIEF PROPAGATION WITH DIFFERENT LLR APPROXIMATIONS}}
\label{table1}
\begin{tabular}[p]{|c|c|c|c|c|c|}
\hline
 \centering
  & $\alpha_A=4.514$ & $\alpha_{\hat{C}}=2.957$ &
  $\alpha_{T}=2.874$ & $\alpha_{opt}=2.957$ & Nonlinear
  Taylor \\ \hline
  $\sigma^{\ast}_z$ & $0.6266$ & $0.6449$ & $0.6445$ & $0.6449$  & $0.6449$ \\
  \hline
  $\frac{E_b}{N_0}^{\ast}$ (dB) & $4.06$ & $3.81$ & $3.82$ & $3.81$ &
  $3.81$ \\ \hline
  \end{tabular}
\end{table*}

\begin{table*}[h]
\center
 \caption{\small{THRESHOLDS FOR THE LDPC ENSEMBLE WITH $\lambda_2(x)= x^3,\:\rho_2(x)=x^{15}$ UNDER
BELIEF PROPAGATION WITH DIFFERENT LLR APPROXIMATIONS}}
\label{table2}
\begin{tabular}[p]{|c|c|c|c|c|c|}
\hline
 \centering
  & $\alpha_A=15.616$ & $\alpha_{\hat{C}}=6.302$ &
  $\alpha_{T}=6.054$ & $\alpha_{opt}=6.287$ & Nonlinear
  Taylor \\ \hline
  $\sigma^{\ast}_z$ & $0.3369$ & $0.3677$ & $0.3674$ & $0.3677$ & $0.3677$ \\
  \hline
  $\frac{E_b}{N_0}^{\ast}$ (dB)& $7.69$ & $6.93$ & $6.94$ & $6.93$ &
  $6.93$ \\ \hline
  \end{tabular}
\end{table*}

\begin{table}[h]
\center
 \caption{\small{LDPC CODES DESIGNED FOR THE UNCORRELATED FLAT RAYLEIGH FADING CHANNEL WITH
UNKNOWN CSI BASED ON APPROXIMATION~(\ref{eqrd}). CODE $1$ and $2$ ARE RATE-
AND THRESHOLD-OPTIMIZED, RESPECTIVELY.}}
\label{table3}
\begin{tabular}[p]{|c|c|c|c|}
\hline
 \centering
& Code $1$ & Code $2$ \\ \hline
$\lambda_2$ & $0.20525$ & $0.20683$ \\
$\lambda_3$ & $0.21067$ & $0.21646$  \\
$\lambda_4$ & $0.00037$ & $0.00046$  \\
$\lambda_5$ &  & $0.00075$ \\
$\lambda_6$ & $0.00180$ & $0.00230$  \\
$\lambda_7$ & $0.18140$ & $0.12574$  \\
$\lambda_8$ & $0.07439$ & $0.13858$ \\
$\lambda_9$ & $0.00248$ & $0.00343$ \\
$\lambda_{10}$ & $0.00099$ & $0.00137$  \\
$\lambda_{11}$ & $0.00059$ & $0.00081$  \\
$\lambda_{15}$ & $0.00026$ & $0.00035$ \\
$\lambda_{20}$ & $0.00024$ & $0.00032$ \\
$\lambda_{29}$ & $0.00181$ & $0.00247$ \\
$\lambda_{30}$ & $0.31975$ & $ 0.30013$ \\
\hline $\rho_9$ & $1.0000$ & $1.0000$  \\
\hline \hline
Rate & $0.4941$ & $0.5000$ \\
$\sigma^{\ast}$ & 0.7436 & 0.7345 \\
${E_b/N_0}^{\ast}\;(dB)$ & $2.63$ & $2.68$ \\ \hline
  \end{tabular}
\end{table}

%\begin{figure}[h]
%\centering
%\includegraphics[width = 0.45\textwidth]{alphaDifference}
%\caption{Comparison of $\alpha_T$ and $\alpha_{\hat{C}}$ for
%different values of $\sigma$.} \label{fig::approxdiff}
%\end{figure}

\begin{thebibliography}{99}

\bibitem{RBA-10}
R. Asvadi, A. H. Banihashemi, and M. Ahmadian-attari,
``Approximation of Log-Likelihood Ratio for Wireless Channels Based
on Taylor Series,'' accepted for presentation at the {\em IEEE
GLOBECOM 2010}, Miami, Florida, Dec. 6 - 10, 2010.

\bibitem{B-76}
R. G. Bartle, {\em The elements of real analysis}, John Wiley \& Sons, 2nd Ed., 1976.

\bibitem{CTB-98}
G. Caire, G. Taricco, and E. Biglieri, ``Bit interleaved coded
modulation,'' \emph{IEEE Trans. Info. Theory}, vol 44, no. 3, pp.
927-946, May 1998.

\bibitem{DD-10}
K. R. Davidson, and A. P. Donsig, {\em Real analysis and
applications-theory in practice}, Springer, 2010.

\bibitem{GL-10}
S. R. Ghorpade, and B. V. Limaye, {\em A course in multivariable
calculus and analysis}, Springer, 2010.

\bibitem{hou}
J. Hou, P. H. Seigel, and L. B. Milestein, ``Performance analysis
and code optimization of low-density parity-check codes on Rayliegh
fading channels,`` \emph{IEEE J. Sel. Areas Comm.}, vol. 19,
no. 5, pp. 924-934, May 2001.

\bibitem{HSMP-03}
J. Hou, P. H. Seigel, L. B. Milestein, and H. D. Pfister,
``Capacity-approaching bandwidth-efficient coded modulation schemes based on low-density
parity-check codes,'' {\em IEEE Trans. Inform. Theory}, vol. 49, no. 9, pp. 2141 - 2155, Sept. 2003.

\bibitem{MB-06}
R. Maddock and A. H. Banihashemi, ``Reliability-Based Coded Modulation with
Low-Density Parity-Check Codes,'' {\em IEEE Trans. Comm.}, vol. 54, no. 3, pp. 403 - 406, March 2006.


\bibitem{rich01B}
T. J. Richardson, M. A. Shokrollahi, and R. L. Urbanke, ``The
capacity of low-density parity-check codes under message-passing
decoding,'' \emph{EEE Trans. Inform. Theory}, vol. 47, no. 2,
pp. 599-618, Feb. 2001.

\bibitem{RSU-01}
T. J. Richardson, M. A. Shokrollahi, and R. L. Urbanke, ``Design of capacity-approaching
irregular low-density parity-check codes,'' \emph{EEE Trans. Inform. Theory}, vol. 47, no. 2,
pp. 619 - 637, Feb. 2001.

\bibitem{SB-07}
H. Saeedi and A. H. Banihashemi, ``Performance of Belief Propagation for Decoding LDPC
Codes in the Presence of Channel Estimation Error,''
{\em IEEE Trans. Comm.}, vol. 55, no. 1, pp. 83 - 89, Jan. 2007.

\bibitem{SB-09}
H. Saeedi and A. H. Banihashemi, ``Design of Irregular LDPC Codes
for BIAWGN Channels with SNR Mismatch,'' {\em IEEE Trans. Comm.},
vol. 57, no. 1, pp. 6 - 11, Jan. 2009.

\bibitem{TB-02}
F. Tosato and P. Bisaglia, ``Simplified soft-output demapper for binary interleaved COFDM
with application to HIPERLAN/2,'' Proc. {\em 2002 IEEE Int. Conf. Comm. (ICC'02)}, 2002, pp. 664 - 668.

\bibitem{YA-09}
R. Yazdani and M. Ardakani, ``Linear LLR approximation for iterative
decoding on wireless channels,'' \emph{IEEE Trans. Comm.}, vol 57,
no. 11, pp. 3278-3287, Nov 2009.

\bibitem{YA-10}
R. Yazdani, and M. Ardakani, ``Efficient LLR calculation for
non-binary modulations over fading channels,'' submitted to
\emph{IEEE Trans. Commun.}, Feb. 2010, [Online]. Available:
arxiv.org/abs/1002.2164

\end{thebibliography}
\end{document}